\documentclass[aps,prb,twocolumn,superscriptaddress,showpacs,footinbib,
               amsfonts,amssymb,amsmath,nobibnotes,reprint,
               ]{revtex4-1}
\usepackage[utf8]{inputenc}
\usepackage{mathrsfs}
\usepackage{graphicx}
\usepackage{color}
\usepackage{braket}
\usepackage{placeins}
\usepackage{pbox}
\usepackage{lipsum}

\usepackage[ps2pdf,extension=pdf,breaklinks]{hyperref}
\usepackage[hyphenbreaks]{breakurl}

\hypersetup{
  colorlinks=true,  % false: boxed links; true: colored links
  linkcolor=blue,   % color of internal links
  citecolor=blue,   % color of links to bibliography
  filecolor=black,  % color of file links
  urlcolor=blue     % color of external links
  }

\newcommand{\tline}{\noalign{\smallskip}\hline\\ \noalign{\vspace{-5pt}}}                                               

\newcommand{\comment}[1]{}

\definecolor{lightgray}{gray}{0.8}
\definecolor{midgray}{gray}{0.5}
\definecolor{darkgray}{gray}{0.3}
\definecolor{darkergray}{gray}{0.2}
\definecolor{darkestgray}{gray}{0.1}

\definecolor{MidPaleAqua}{RGB}{63,128,155}
\definecolor{EmpireGreen}{RGB}{39,81,42}
\definecolor{RoyalBlue}{RGB}{29, 56,126}
\definecolor{VertDeGris}{RGB}{59,117,71}
\definecolor{Tomette}{RGB}{155,65,49}
\definecolor{MidAqua}{RGB}{50,119,178}

\let\oldlipsum\lipsum                                                            
\renewcommand{\lipsum}[1][1]{{\color{lightgray}{\oldlipsum[#1]}}}

\newcommand{\mv}[1]{\mathbf{#1}}
\newcommand{\Tr}{\text{Tr}}
\newcommand{\sign}{\text{sgn}}

\renewcommand{\Im}{\mathfrak{Im}}
\renewcommand{\Re}{\mathfrak{Re}}

\newcommand{\ieig}{\lambda}
\newcommand{\jeig}{\lambda'}
\newcommand{\keig}{\lambda''}

\newcommand{\feig}{f_{\ieig}}
\newcommand{\eig}{\varepsilon_{\ieig}}
\newcommand{\eigi}{\varepsilon_{\ieig}}

\newcommand{\eigoi}{\varepsilon^0_{\ieig}}
\newcommand{\eigoj}{\varepsilon^0_{\jeig}}
\newcommand{\eigok}{\varepsilon^0_{\keig}}
\newcommand{\eigo}{\eig^0}

\newcommand{\wfni}{\psi_{\ieig}}
\newcommand{\wfnj}{\psi_{\jeig}}
\newcommand{\wfnk}{\psi_{\keig}}

\newcommand{\icell}{l}
\newcommand{\iat}{\kappa}
\newcommand{\icart}{j}

\renewcommand{\vr}{\mv{r}}
\newcommand{\vR}{\mv{R}}
\newcommand{\vq}{\mv{q}}

\newcommand{\vk}{\mv{k}}

\newcommand{\ipho}{\nu}
\newcommand{\jpho}{\nu'}

\newcommand{\freqi}{\omega_{\ipho}}
\newcommand{\freqj}{\omega_{\jpho}}

\newcommand{\one}{^{\tiny (1)}}

\newcommand{\ones}{^{\tiny (1)*}}
\newcommand{\two}{^{\tiny (2)}}

\newcommand{\del}{\partial}

\newcommand{\idallz}{\int d \allz}
\newcommand{\allz}{\mathbf{z}}
\newcommand{\alln}{\mathbf{n}}
\newcommand{\normallchi}{\big\vert \chi^{\alln}(\allz) \big\vert^2 }

\newcommand{\etal}{{\it et al. }}

\newcommand{\udem}{D\'{e}partement de physique, Universit\'{e} de Montr\'{e}al,
                   C.P. 6128, Succursale Centre-Ville, Montr\'{e}al,
                   Canada H3C 3J7}
\newcommand{\ucl}{Institute of Condensed Matter and Nanosciences,
                  Universit\'{e} catholique de Louvain,
                  B-1348 Louvain-la-Neuve, Belgium}

\begin{document}

\pacs{63.20.kd, 63.20.dk, 78.20.-e, 71.15.Mb, 71.20.Nr}
\title{Dynamical and anharmonic effects on the electron-phonon coupling
       and the zero-point renormalization of the electronic structure}
\author{G. Antonius} \email{gabriel.antonius@gmail.com} \affiliation{\udem}
\author{S. Ponc\'e}   \affiliation{\ucl}
\author{E. Lantagne-Hurtubise}  \affiliation{\udem}
\author{G. Auclair}   \affiliation{\udem}
\author{X. Gonze}     \affiliation{\ucl}
\author{M. C\^ot\'e}  \affiliation{\udem}

\begin{abstract}

The renormalization of the band structure at zero temperature
  due to electron-phonon coupling is explored
  in diamond, BN, LiF and MgO crystals.
We implement a dynamical scheme to compute the frequency-dependent self-energy
  and the resulting quasiparticle electronic structure.
Our calculations reveal the presence of a satellite band
  below the Fermi level of LiF and MgO.
We show that the renormalization factor (Z),
  which is neglected in the adiabatic approximation,
  can reduce the zero-point renormalization (ZPR)
  by as much as $40\%$.
Anharmonic effects in the renormalized eigenvalues
  at finite atomic displacements are explored with the frozen-phonon method.
We use a non-perturbative expression for the ZPR,
  going beyond the Allen-Heine-Cardona theory.
Our results indicate that high-order electron-phonon coupling terms
  contribute significantly to the zero-point renormalization
  for certain materials.

\end{abstract}

\maketitle

The electron-phonon coupling is at the heart of numerous phenomena
  such as optical absorption
  \cite{Kioupakis2010,Noffsinger2012},
  thermoelectric transport \cite{Wang2011},
  and superconductivity \cite{Cardona2006,Giustino2007a,Lee2004,Blase2004}.
It is also a crucial ingredient in basic electronic structure calculations,
  giving renormalized quasiparticle energies and lifetimes.
This renormalization causes the temperature dependence of the band gap
  of semiconductors \cite{Cardona2005}, 
  and accounts for the zero-point renormalization (ZPR),
  while the lifetime broadenings are observed through the electron mobility
  \cite{Restrepo2009,Tyuterev2011}
  and in photo-absorption/emission experiments
  \cite{Bhattacharya2015}.

Obtaining the quasiparticle structure from first principles
  has been a challenge, addressed for the first time for bulk silicon
  by King-Smith \etal \cite{King-Smith1989}, in 1989,
  using density functional theory (DFT),
  with a mixed frozen-phonon supercell and linear response approach.
These authors pointed the inadequate convergence of their results with respect
  to phonon wavevector sampling,
  due to the limited available computing capabilities.
Fifteen year passed,
  before Capaz \etal computed it for carbon nanotubes \cite{Capaz2005}
  using DFT with frozen phonons.
At variance with the frozen-phonon approach, the theory of Allen, Heine
  and Cardona (AHC) \cite{Allen1976,Allen1981,Allen1983}
  casts the renormalization and the broadening in terms of the first-order
  derivatives of the effective potential with respect to atomic positions.
Used initially with empirical potentials, tight-binding or semi-empirical
  pseudopotentials \cite{Allen1976, Allen1981,Allen1983,
  Zollner1991,Zollner1992,Garro1996,Olguin2002}, AHC was then implemented
  with the density functional perturbation theory (DFPT)
  \cite{Baroni1987,Baroni2001,Gonze1997,Gonze1997a},
  providing an efficient way to compute the phonon band structure
  and the electron-phonon coupling altogether.
This powerful technique allowed A. Marini to compute, from first principles,
  temperature-dependent optical properties \cite{Marini2008}.

While DFPT has been widely applied to predict structural and thermodynamical
  properties of solids \footnote{
  See for example some of the largely cited article introducing DFPT
  \cite{Baroni1987,Gonze1997}.
  },
  few studies have used it to compute
  the phonon-induced renormalization of the band structure.
The scarcity of experimental data is at least partly responsible
  for this imbalance.
Whereas the phonon spectrum is commonly measured through Raman spectroscopy
  and neutron-scattering experiments,
  evaluating the ZPR requires low-temperature ellipsometry measurements
  or isotope substitutions, which are less abundant in literature.
From a theoretical point of view, the calculation of the ZPR
  relies on several assumptions that we will be addressing in this paper.

We identify two kinds of approximations.
The first kind are those regarding the treatment
  of the electron-electron interactions, which is achieved in DFT
  through the Hartree and the exchange-correlation potentials.
It was shown that the strength of the electron-phonon interaction
  was highly sensitive to the choice of the exchange-correlation functional
  \cite{LaflammeJanssen2010}.
Subsequent GW calculations confirmed that standard functionals such as
  the local-density approximation (LDA) tend to underestimate
  the electron-phonon coupling by as much as 30\%
  \cite{Faber2011,Lazzeri2008,Yin2013,Antonius2014}.

The second kind of approximations are those made on the self-energy
  of the electron-phonon interaction.
One, for example, usually performs the \emph{rigid-ion approximation},
  assuming that the second-order derivative of the Hamiltonian
  is diagonal in atom sites.
This approximation proved to be valid in the case of crystals
  \cite{Antonius2014,Ponce2014a},
  but notably fails for diatomic molecules \cite{Gonze2011}.

Another assumption is the \emph{adiabatic approximation}, which implies that
  the phonon population can be treated as a static perturbation.
One would typically compute the real part of the self-energy
  in a static way, and use a dynamical expression to compute
  the imaginary part and obtain the electronic lifetimes \cite{Giustino2010}.
The adiabatic approximation breaks downs in certain materials
  such as diamond and polyacetylene,
  as pointed out by Cannuccia and Marini \cite{Cannuccia2011,Cannuccia2012}.
By considering the frequency dependence of the self-energy, they showed that
  the electron-phonon interaction smears out the energy levels,
  even obliterating the band structure.

Finally, the \emph{harmonic approximation} is the assumption that the
  total energy and electronic eigenvalues vary quadratically with
  atomic displacements, which justifies the use of a
  second-order perturbation theory.
Higher order expansions have been used to compute phonon
  wavefunctions, energies, and thermal expansion coefficients
  \cite{Rignanese1996,Monserrat2013, Paulatto2013},
  but its impact on the ZPR was rarely investigated. 

In this work, we compute the ZPR and the quasiparticle lifetimes
  of the band structure of diamond, BN, LiF and MgO.
We show that the inclusion of dynamical effects in the AHC theory
  is important to obtain correct quasiparticle energies and broadenings.
We also study the impact of anharmonic effects in the electronic energies
  by means of frozen-phonon calculations,
  and show that high-order terms do contribute
  to the electron-phonon coupling in certain cases.

All calculations are performed with the {\sc Abinit} code \cite{Gonze2009}.
For simplicity, we use an LDA exchange-correlation functional.
We do not expect this approach to fully capture the strength of the coupling,
  as does the GW method.
Rather, it allows us to evaluate the impact of several commonly adopted
  approximations to the electron-phonon coupling self-energy.

\section{Dynamical DFPT}

The dynamical AHC theory is derived by expanding
  the starting Hamiltonian $H_0$
  up to second order in atomic displacements as (using atomic units)
\begin{align} \label{Hep}
  H_{ep} = H_0 &
    + \sum_{\ieig \jeig \ipho}
        \frac{1}{\sqrt{2N\freqi}} \Bra{\wfni} V\one_{\ipho} \Ket{\wfnj}
        \ A_{\ipho} c^{\dagger}_{\ieig} c_{\jeig} \nonumber\\
   + & \sum_{\ieig \jeig \ipho \jpho}
        \frac{1}{2N\sqrt{\freqi\freqj}}
        \Bra{\wfni} V\two_{\ipho\jpho} \Ket{\wfnj}
        \ A_{\ipho} A_{\jpho} c^{\dagger}_{\ieig} c_{\jeig}
,
\end{align}
where $c^{\dagger}_{\ieig}$ and $c_{\ieig}$ are electron
  creation and destruction operators,
  and $A_{\ipho} = a_{\ipho} + a^{\dagger}_{-\ipho}$, such that
  $A_{\ipho}/\sqrt{2\freqi}$ represents a phonon displacement operator.
The electronic states $\ieig$ with wavefunctions $\wfni$ and energies $\eigoi$
  are specified by a wavevector $\vk_{\ieig}$, a band index $b_{\ieig}$,
  and spin $\sigma_{\ieig}$,
  while the phonon modes $\ipho$ with frequencies $\freqi$
  are specified by a wavevector $\vq_{\ipho}$ and a branch index $m_{\ipho}$,
  and $N$ is the number of phonon wavevectors.

The first-order perturbation potential is formed
  with derivatives of the Hamiltonian
  with respect to atomic displacements along a particular phonon mode as
\begin{equation}
  V\one_{\ipho} =
     \nabla_{\ipho} \ H_0 =
     \sum_{\icell\iat\icart}
      U^{\ipho}_{\iat\icart}
      e^{i \vq_{\ipho} \cdot \vR_{\icell}}
      \nabla_{\icell\iat\icart} \ H_0
,
\end{equation}
where $\icell$ labels a unit cell with lattice vector $\vR_{\icell}$,
  $\iat$ an atom within a unit cell,
  $\icart$ a cartesian direction,
  and $U^{\ipho}_{\iat\icart}$ is the phonon displacement vector.
The second-order perturbation potential is then
${V\two_{\ipho\jpho}(\vr) = \tfrac{1}{2} \nabla_{\ipho} \nabla_{\jpho}^* H_0}$.

\smallskip

Within the DFT and DFPT approaches of this work,
  $H_0$ is an electron-only Hamiltonian, and
  the phonon perturbations involve the derivatives of
  the local self-consistent potential.
Since the electronic density responds to the atomic displacements
  by screening the ions' potential,
  these perturbations must be evaluated self-consistently.
Furthermore, the phonon displacement vectors are {\it a priori} unknown
  and must be computed by diagonalization of the the dynamical matrix.
Other formulations of $H_0$ could include the many-body interactions 
  between the electrons and the ions~\cite{marini_manybody_2015}.
These alternatives however fall beyond the scope of this work.

\smallskip
\smallskip

Following the usual many-body perturbation theory \cite{Mahan2000},
  the electron-phonon self-energy at second order
  is the sum of the Fan and the Debye-Waller terms:
\begin{equation}
  \Sigma^{ep}_{\ieig \jeig}(\omega) =
    \ \Sigma^{Fan}_{\ieig \jeig}(\omega)
  \ + \ \Sigma^{DW}_{\ieig \jeig}
.
\end{equation}
The dynamical Fan term is given by
\begin{widetext}
\begin{align} \label{FanSE}
   \Sigma^{Fan}_{\ieig\jeig}(\omega) =
    \sum_{\ipho} \frac{1}{2\freqi}
    \sum_{\keig}
    \Bra{\wfni} V\one_{\ipho} \Ket{\wfnk}
    \Bra{\wfnk} V\ones_{\ipho} \Ket{\wfnj}
    \bigg[ \frac{n_\ipho(T) + f_{\keig}(T)}
        {\omega - \eigok + \freqi + i\eta \ \sign(\omega)}
        + \frac{n_\ipho(T) + 1 - f_{\keig}(T)}
        {\omega - \eigok - \freqi + i\eta \ \sign(\omega)}
    \bigg]
,
\end{align}
\end{widetext}
where $n_{\ipho}$ and $f_{\ieig}$ are boson and fermion
  occupation factors, and $\eta$ is a small parameter
  which is real and positive.
This parameter maintains causality by giving the correct sign to the imaginary
  part of the quasiparticle energies.
It also smooths the frequency dependence of the self-energy
  when a finite sampling of phonon modes is used (see appendix~\ref{app:eta}).
We note that the periodicity of the phonon perturbation potential
  restricts the summation over intermediate states
  to those at the k-point given by
  ${\vk_{\keig}+\vq_{\ipho}=\vk_{\ieig}=\vk_{\jeig}}$.
In Eq.~\eqref{FanSE} and in the remaining of this work,
  all the summations over the phonon modes
  are implicitly normalized by the number of wavevectors
  used to sample the Brillouin zone.

The frequency-independent Debye-Waller term is formally defined as
\begin{equation} \label{DWSE}
  \Sigma^{DW}_{\ieig \jeig} =
    \sum_{\ipho} \frac{1}{2\freqi}
    \Bra{\wfni} V\two_{\ipho\ipho} \Ket{\wfnj} \big[2n_{\ipho}(T) + 1\big]
,
\end{equation}
which also implies $\vk_{\ieig}=\vk_{\jeig}$.
Within the rigid-ion approximation,
  the Debye-Waller term can be computed
  using only the matrix elements of $V\one_{\ipho}$,
  in a form similar to the Fan term,
  thanks to translational invariance \cite{Ponce2014}.

The interacting Green's function is the solution of the Dyson equation
  involving the full electron-phonon self-energy,
  which is diagonal in $(\vk_{\ieig},\vk_{\jeig})$.
If the bands are well separated in energy, then the Green's function
  %at a given k-point
  can be approximated with the diagonal elements of the self-energy as
\begin{equation}
  G_{\ieig}(\omega) \approx
    \big(\omega - \eigo - \Sigma^{ep}_{\ieig}(\omega) \big)^{-1}
,
\end{equation}
where we use the shorthand
  $\Sigma^{ep}_{\ieig}(\omega)\equiv\Sigma^{ep}_{\ieig\ieig}(\omega)$.
By considering only the diagonal elements of the self-energy,
  we disregard the possible change in the electronic density
  resulting from the electron-phonon interaction.
Taking this effect into account would involve
  solving the Dyson equation for the Green's function
  to obtain a new density,
  and applying the change in the DFT self-consistent potential perturbatively.
We are assuming however that the change in the one-electron state densities
  among the set of occupied bands (and among the set of unoccupied bands)
  would compensate, and that the additional perturbative terms
  (that is, excitations accross the gap) be negligible.
Besides, we stress that
  the Green's function only needs to be corrected once,
  since this procedure is a second-order perturbative approach.
Any attempt for self-consistency in the calculation of the Green's function
  and the self-energy belongs to a higher-order treatment.

\begin{figure}[ht]
\includegraphics[width=1.0\linewidth]{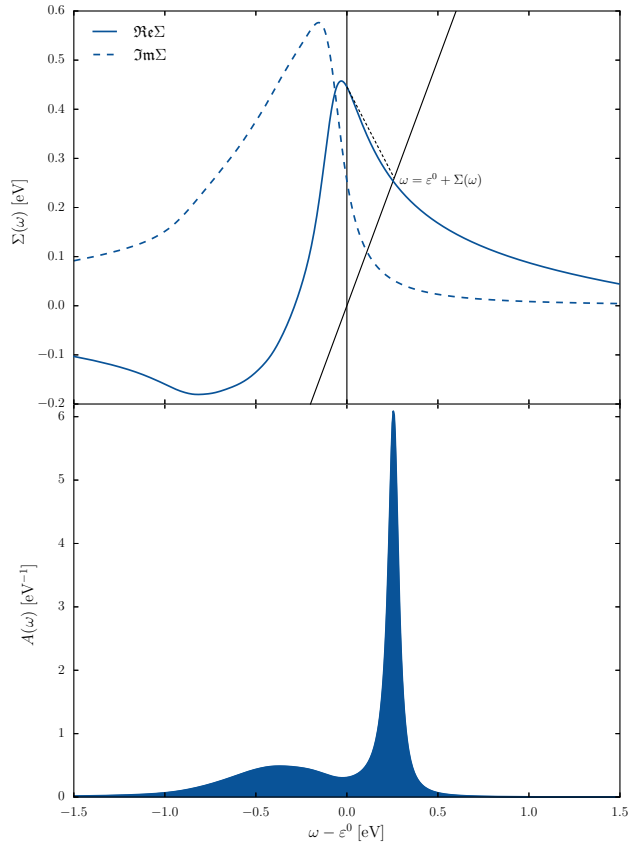}
  \caption{\label{fig:SelfEnergySpectralLiF}
    Upper: Real and imaginary part of the self-energy
      for the top of the valence bands (VB) of LiF.
    The vertical line indicates the bare eigenvalue,
      the $x=y$ line gives the renormalized eigenvalue
      at the intersection of the real part of the self-energy,
      and the short-dashed line is the linearized self-energy,
      which approximates the renormalized eigenvalue at the intersection
      of the $x=y$ line. 
    Lower: The corresponding spectral function.
    The position of the principal peak gives the quasiparticle energy.
    This narrow peak collects $\sim60\%$ of the weight,
      and the rest of the charge forms a broad satellite peak below
      the bare eigenvalue.
    Note that the finite width of this peak is an artifact of the
      imaginary parameter used ($0.1$~eV).
           }
\end{figure}

From the imaginary part of the Green's function,
  one obtains the spectral function
\begin{equation}
  A_{\ieig}(\omega)
    = \frac{1}{\pi} \frac{\vert \Im \Sigma^{ep}_{\ieig}(\omega)\vert}
      {[\omega - \eigo - \Re \Sigma^{ep}_{\ieig}(\omega)]^2
       + \Im \Sigma^{ep}_{\ieig}(\omega)^2}
,
\end{equation}
which directly relates to the signal observed in ARPES experiments.
The quasiparticle energies $\eig$ are defined as the position of the principal peak
  of $A_{\ieig}(\omega)$.
Neglecting the frequency dependence of the imaginary part of the
  self-energy, the maximum of the spectral function is at
  the energy given by
\begin{align} \label{QPenergy}
  \eig = \eigo + \Re \Sigma^{ep}_{\ieig}(\eig)
.
\end{align}
Assuming furthermore that the quasiparticle energies
  are close to the bare electronic energies,
  the latter can be corrected perturbatively as
\begin{equation}
  \eig \approx \eigo + Z_{\ieig} \Re \Sigma^{ep}_{\ieig}(\eigo)
\end{equation}
where
\begin{equation}
Z_{\ieig} = \Big(1 - \Re \frac{\partial \Sigma^{ep}_{\ieig}(\omega)}
              {\partial \omega} \big\vert_{\omega=\eigoi}\Big)^{-1}
\end{equation}
is the renormalization factor.
This procedure accounts for a linearization of the self-energy
  near the bare eigenvalue,
  as illustrated in Fig~\ref{fig:SelfEnergySpectralLiF}.

The quasiparticle broadening $\gamma_{\ieig}$
  is defined as the half width of the spectral function at half of its maximum,
  which means for a symmetrical quasiparticle peak that
  ${A_{\ieig}(\eig \pm \gamma_{\ieig}) = {A_{\ieig}(\eig)}/{2}}$.
Neglecting the frequency dependence of the self-energy
  near the quasiparticle energies, the broadening can be approximated
  as $\vert \Im \Sigma^{ep}_{\ieig}(\eigi) \vert$,
  the imaginary part of the self-energy, which writes
\begin{align} \label{ImFanSE}
   \vert \Im \Sigma^{ep}_{\ieig}(\omega) \vert = 
    \sum_{\ipho} \frac{1}{2\freqi}
    \sum_{\jeig} & 
    \vert \Bra{\wfni} V\one_{\ipho} \Ket{\wfnj} \vert^2\nonumber\\
   \times \big[ \big( n_{\ipho} + f_{\jeig}\big)
     &  \delta(\omega - \eigoj + \freqi)\nonumber\\
    + \big(n_{\ipho} + 1 - f_{\jeig}\big)
      & \delta(\omega - \eigoj - \freqi)
    \big]
.
\end{align}

One recovers the static AHC expression for the ZPR and the broadening
  by neglecting the phonon frequencies in the self-energy.
Such approximation is made on the basis that the phonon frequencies
  and the quasiparticle corrections are much smaller
  than the typical energy differences with transition states
  that contribute to the self-energy.
Under this assumption, the static Fan term reads
\begin{align} \label{staticFanSE}
  \Sigma^{\substack{static\\Fan}}_{\ieig} =
    \sum_{\ipho} \frac{1}{2\freqi}
  \sum_{\jeig} &
    \frac{
    \vert \Bra{\wfni} V\one_{\ipho} \Ket{\wfnj} \vert^2 
    }{\eigoi - \eigoj + i \eta \ \sign(\eigoi)}
    \big[ 2 n_\ipho(T) + 1 \big]
.
\end{align}
Evaluating Eq.~\eqref{staticFanSE} instead of Eq.~\eqref{FanSE}
  requires much less computational efforts,
  especially if the Sternheimer equation is used
  to eliminate the summation over high-energy electron bands.

In this work, we adopt a semi-static approximation to compute
  the frequency-dependent self-energy.
The terms of Eq.~\eqref{FanSE} are being computed explicitly,
  up to a certain band index $b_{\ieig}^{\rm max}$,
  and the contribution of the remaining bands above $b_{\ieig}^{\rm max}$
  is treated statically with Eq.~\eqref{staticFanSE}
  using the Sternheimer equation method \cite{Gonze2011}.
For our materials, the bands above $b_{\ieig}^{\rm max}$
  lie more than $20$~eV above the states being corrected.
Hence, the relative error on the self-energy due to the static treatment
  of these bands' contributions is a few percent at most.

\subsection*{Results and discussion}

We compute the quasiparticle structure of four crystalline materials:
  diamond (C), boron nitride (BN) in the zinc-blende structure,
  magnesium oxide (MgO), and lithium fluoride (LiF)
  in the rock-salt structure.
For all materials, we use a $8\times8\times8$ k-point grid
  for the electronic wavefunctions and density,
  and a $32\times32\times32$ q-point grid for the phonon modes sampling.

\begin{figure*}
\includegraphics[width=0.49\linewidth]{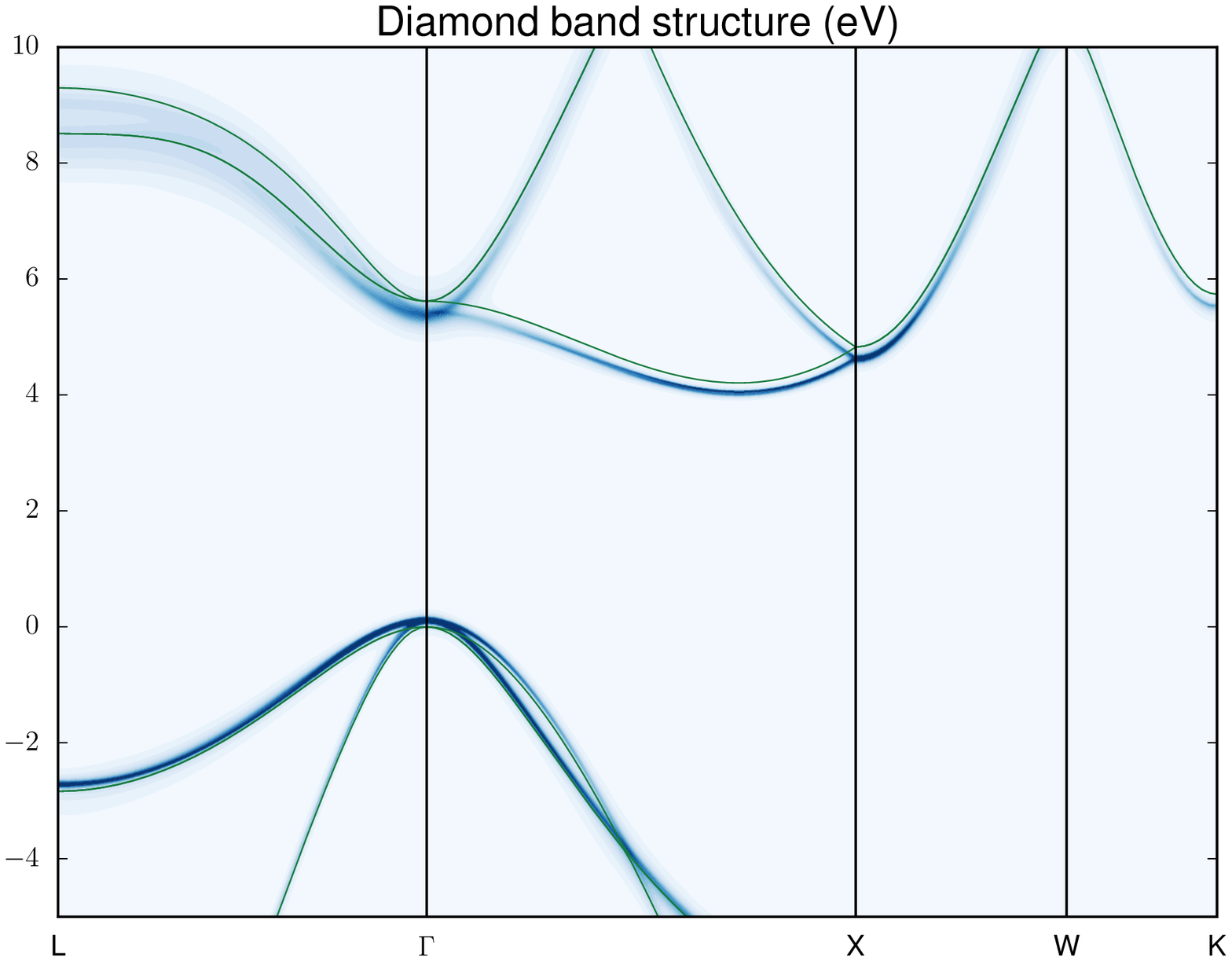}
\includegraphics[width=0.49\linewidth]{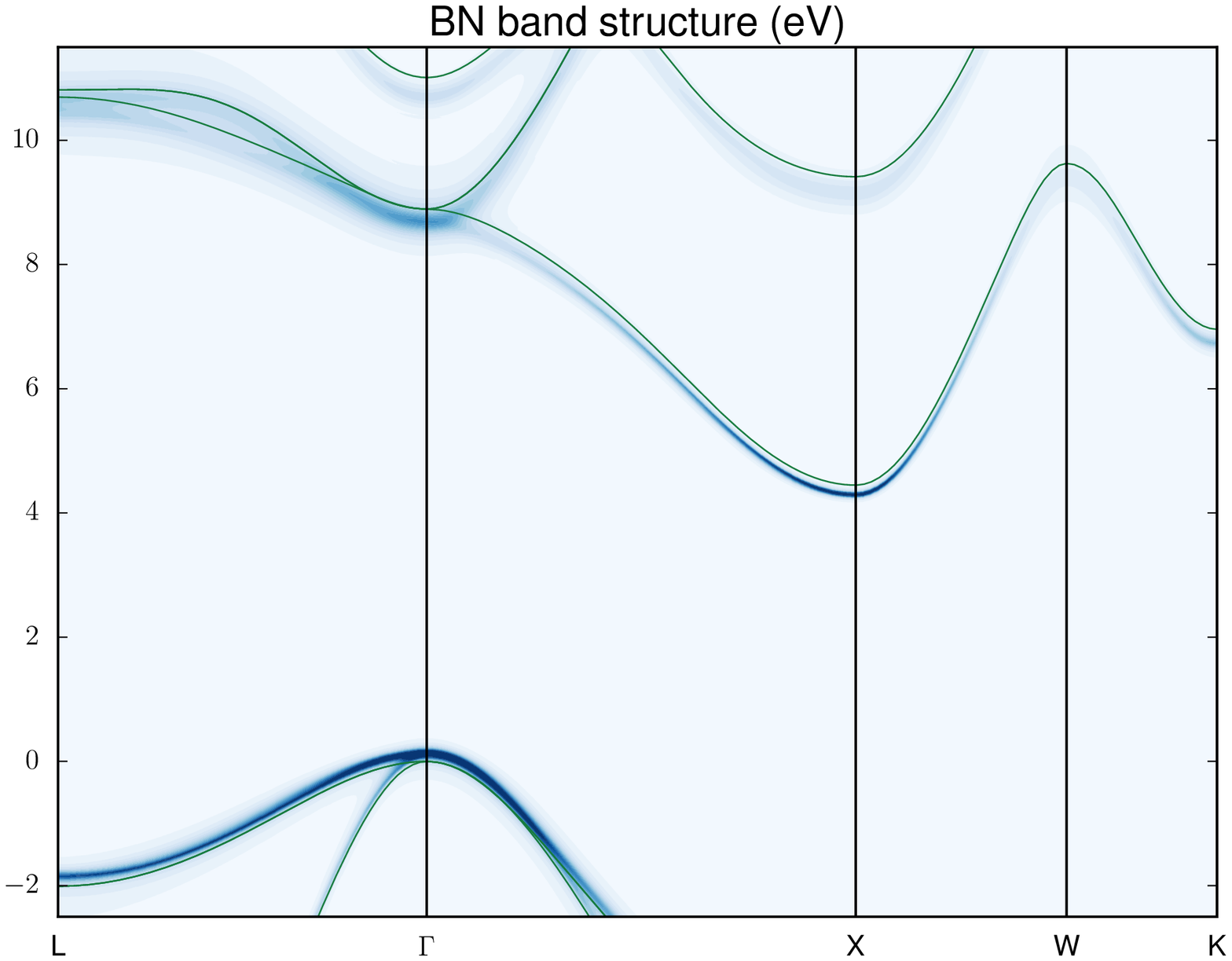}\\
\includegraphics[width=0.49\linewidth]{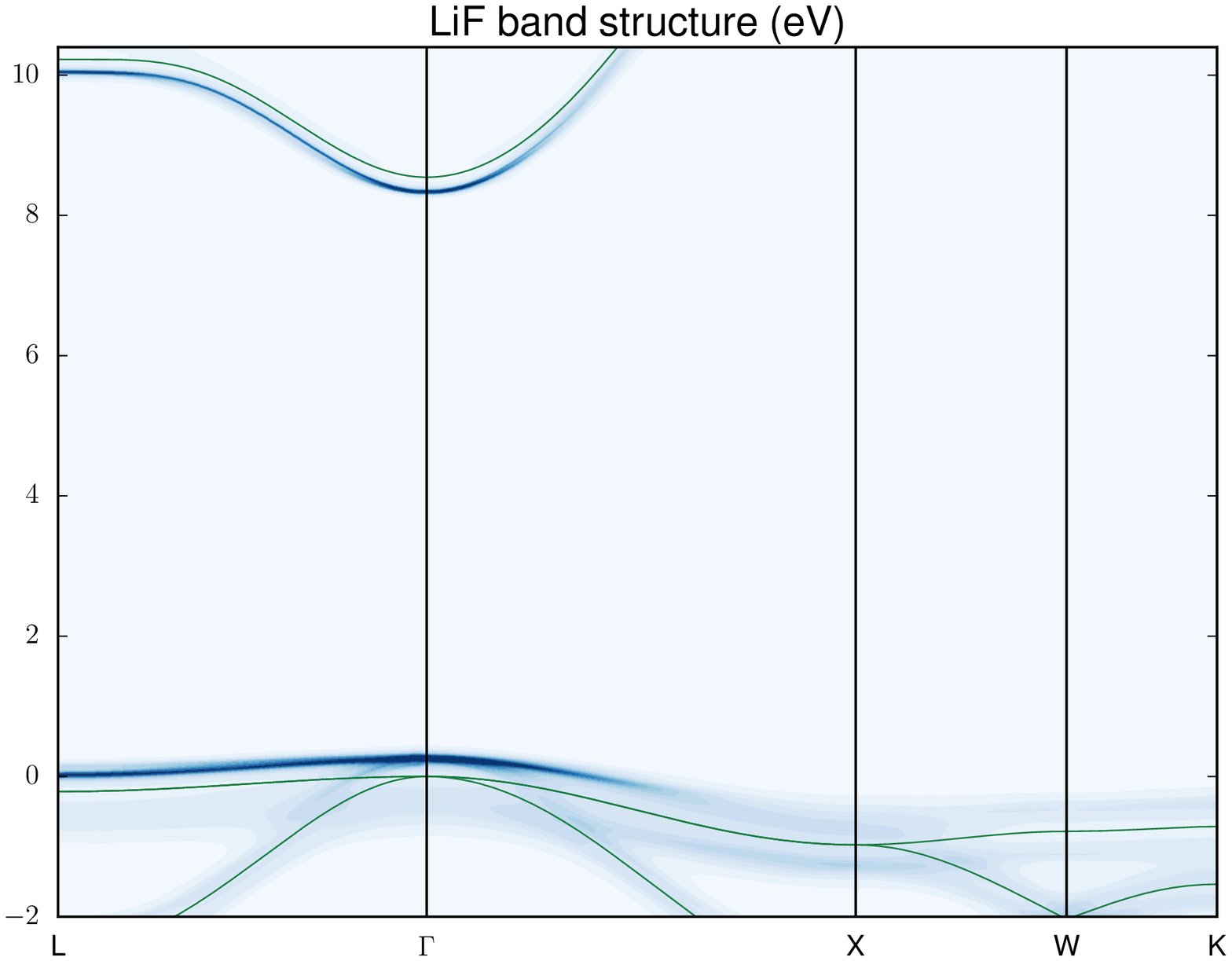}
\includegraphics[width=0.49\linewidth]{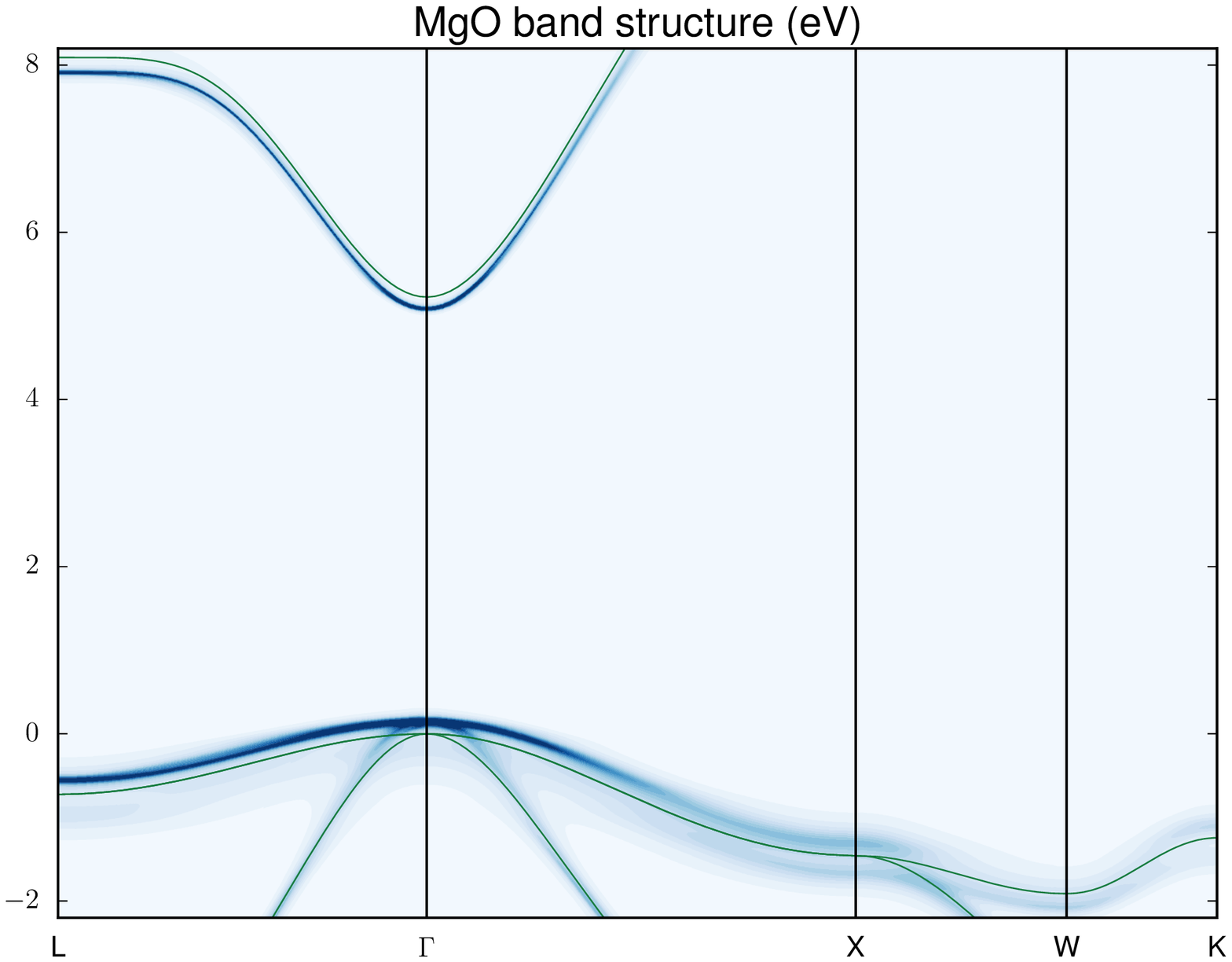}\\
  \caption{\label{fig:BandStructureSpectralFunction}
    Spectral functions summed over the bands at each k-point of
      the Brillouin zone (arbitrary units).
    The green lines are the DFT band structure, in eV.
    When a quasiparticle peak is visible in the spectral function,
      the renormalization is infered from the difference of the position
      of that peak with the bare band structure. 
    In the regions of flat bands, the band structure is being completely
      diffused.
    A satellite peak is seen below the last valence band of LiF and MgO.
           }
\end{figure*}

The spectral functions at zero temperature
  are shown in Fig.~\ref{fig:BandStructureSpectralFunction}
  for the full band structure.
A distinctive quasiparticle peak appears at the band edges,
  shifted from the bare eigenvalues,
  while in the regions of flat bands
  the spectral function is being diffused.
The first conduction band of the indirect band gap materials (diamond, BN)
  exhibits a strong renormalization and a large broadening at $\Gamma$.
This is due to the presence of other states in the Brillouin zone
  with close energies that are available for scattering.

Another striking feature is the last valence band of LiF and MgO
  being completely diffused due to strong intra-band coupling.
They show a narrow quasiparticle peak above the bare eigenvalue,
  and a broad satellite peak below.
These features originate from the frequency dependence of both
  the real and the imaginary parts of the self-energy,
  as shown in Fig.~\ref{fig:SelfEnergySpectralLiF}
  for the top of the valence bands of LiF.
The satellite peaks could be observed in ARPES measurements,
  such as those performed on MgO by Tjeng \etal \cite{tjeng_electronic_1990}.
However, a direct comparison with our results would require
  the full experimental spectra along the high-symmetry lines.

Table~\ref{tab:dynzpr} presents the real part of the self-energy
  for the states forming the optical band gap,
  namely the top of the valence bands (VB),
  and the first conduction band (CB) at $\Gamma$.
At the bare eigenvalues, the self-energy shows little difference
  between the static and dynamical DFPT schemes,
  indicating that the phonon frequencies could be safely ignored
  in its real part.
However, the frequency dependence produces an important
  renormalization factor $Z$,
  ranging from $0.60$ to $0.93$ for the valence bands,
  and from $0.75$ to $1.0$ for the conduction bands.
Thus, the dynamical effects tend to reduce the zero-point correction,
  with respect to the static scheme.
Comparing the linearized self-energy with the quasiparticle correction
  obtained by solving Eq.~\eqref{QPenergy} numerically
  or from the position of the principal peak of the spectral function,
  the linearization scheme proves to be a good approximation to both.

The renormalization factor being larger than~$1$
  indicates a breakdown of the quasiparticle picture.
If the imaginary part of the self-energy is small,
  there is a well defined quasiparticle peak,
  and $Z$ can be interpreted as the weight of that peak,
  which has to be smaller than~$1$.
Otherwise, the spectral function is diffused,
  there is no such interpretation for $Z$ and its value is unconstrained.
  %and the value of $Z$ is unconstrained.

Table~\ref{tab:broadening} presents
  the quasiparticle broadening
  of the indirect-band gap materials
  computed with various schemes.
The difference in the broadenings obtained from
  the static and the dynamical DFPT schemes
  is best understood with Eq.~\eqref{ImFanSE}.
Only the electronic states in a narrow energy range
  are available for scattering.
The imaginary part of the self-energy is thus sensitive
  to the inclusion of phonon frequencies,
  since they affect the positioning of this energy range.
For the same reason, the broadening varies rapidly with frequency,
  which results in an important difference
  between the imaginary part at the \emph{bare} eigenvalue
  and that at the \emph{renormalized} eigenvalue.
Comparing these values with the actual width of the quasiparticle peak,
  we conclude that only the imaginary part of the self-energy evaluated
  at the renormalized energy is an accurate estimation of the broadening.

\begin{table*}
  \caption{
    Zero-point renormalization (in eV) evaluated
      from the real part of the self-energy 
      using a static expression ($stat$), a dynamical expression ($dyn$),
      at the bare eigenvalue ($\varepsilon^0$),
      at the renormalized eigenvalue ($\varepsilon$),
      or from the displacement of the main quasiparticle peak
      from the bare eigenvalue ($\Delta A(\varepsilon)$).
    The unitless renormalization factor $Z$
      is used to linearize the self-energy near the bare eigenvalue.
    See Apprendix~\ref{app:eta} for the values of the $\eta$ parameter used.
    \label{tab:dynzpr}}
  \begin{ruledtabular}
  \begin{tabular}{l l | r r r r r r}
    & &
    $\Sigma^{stat}(\varepsilon^0)$ &
    $\Sigma^{dyn}(\varepsilon^0)$ &
    $Z$ &
    $Z \Sigma^{dyn}(\varepsilon^0)$ &
    $\Sigma^{dyn}(\varepsilon)$ &
    $\Delta A(\varepsilon)$\\
    \hline
    C &    VB &   0.134 &   0.126 &   0.931 &   0.118 &   0.118 &   0.118\\
      &    CB &  -0.238 &  -0.240 &   1.007 &  -0.242 &  -0.240 &  -0.247\\
      &   Gap &  -0.372 &  -0.366 &         &  -0.359 &  -0.358 &  -0.365\\
   BN &    VB &   0.198 &   0.173 &   0.823 &   0.143 &   0.147 &   0.147\\
      &    CB &  -0.190 &  -0.196 &   1.020 &  -0.200 &  -0.197 &  -0.208\\
      &   Gap &  -0.388 &  -0.370 &         &  -0.343 &  -0.344 &  -0.355\\
  MgO &    VB &   0.197 &   0.198 &   0.734 &   0.145 &   0.145 &   0.147\\
      &    CB &  -0.153 &  -0.143 &   0.870 &  -0.125 &  -0.127 &  -0.127\\
      &   Gap &  -0.350 &  -0.341 &         &  -0.270 &  -0.272 &  -0.274\\
  LiF &    VB &   0.398 &   0.446 &   0.596 &   0.266 &   0.254 &   0.256\\
      &    CB &  -0.279 &  -0.273 &   0.746 &  -0.204 &  -0.211 &  -0.211\\
      &   Gap &  -0.677 &  -0.718 &         &  -0.469 &  -0.464 &  -0.467\\
  \end{tabular}
  \end{ruledtabular}
\end{table*}

\begin{table}
  \caption{
    Quasiparticle broadening (in eV) evaluated
      from the imaginary part of the self-energy
      using a static expression ($stat$), a dynamical expression ($dyn$),
      at the bare eigenvalue ($\varepsilon^0$),
      at the renormalized eigenvalue ($\varepsilon$),
      or from the width of the main quasiparticle peak at half of its maximum
      ($\gamma$).
    See Apprendix~\ref{app:eta} for the values of the $\eta$ parameter used.
    \label{tab:broadening}}
  \begin{ruledtabular}
  \begin{tabular}{l l | c c c c}
    & &
    $\vert \Im \Sigma^{stat}(\varepsilon^0)\vert$ &
    $\vert \Im \Sigma^{dyn}(\varepsilon^0) \vert$ &
    $\vert \Im \Sigma^{dyn}(\varepsilon) \vert$ &
    $\gamma$\\
    \hline
    C &    CB &   0.178 &   0.164 &   0.140 &   0.138\\
   BN &    CB &   0.246 &   0.226 &   0.200 &   0.196\\
  \end{tabular}
  \end{ruledtabular}
\end{table}

\section{Anharmonic effects}

\begin{figure}
  \includegraphics[width=1.0\linewidth]{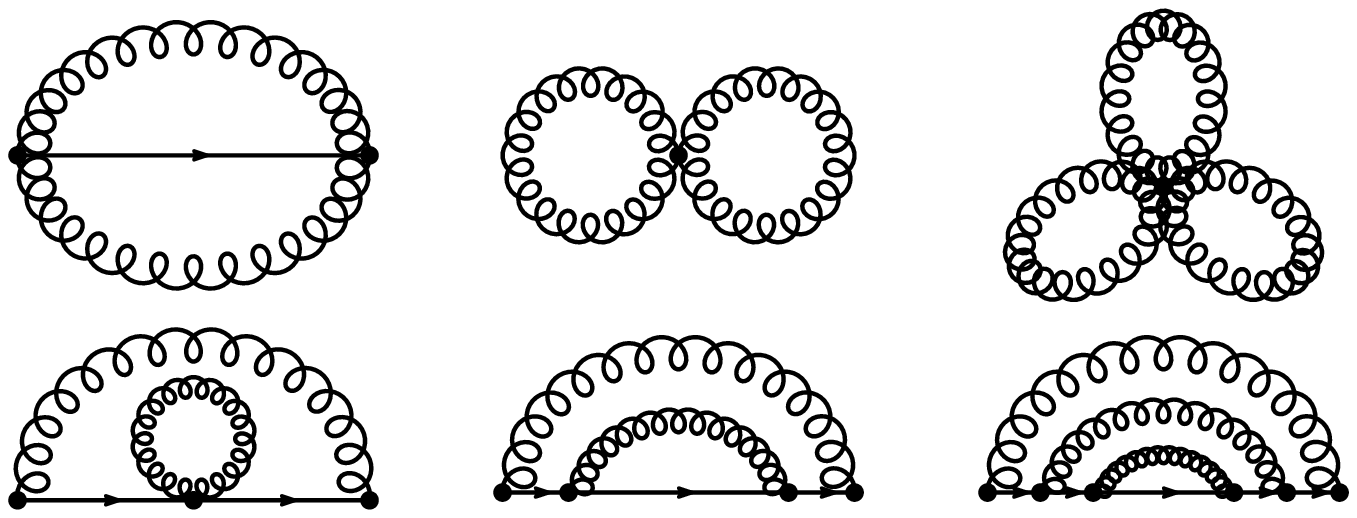}
  \caption{\label{highorder}
    Some of the high-order electron-phonon coupling diagrams
      which contribute to the self-energy.
    Each vertex with $n$ phonon branches is associated with
      a $n^{th}$-order derivative of the one-particle Hamiltonian.
    The independent-phonon approximation presented in the text
      retains only the diagrams formed with multiple interactions
      involving the \emph{same} phonon mode. 
    }
\end{figure}

The frozen-phonon method allows for a direct computation of the
  electron-phonon self-energy within the adiabatic approximation.
We present here an extension of this method,
  which allows to explore anharmonic effects
  beyond the second-order perturbation theory of Allen, Heine and Cardona.

Recalling the theory of the harmonic crystal,
  we write the total energy in a frozen-phonon configuration as
\begin{equation}
  E[\allz] = E_0 + \sum_{\ipho} \frac{\freqi^2}{2} z_{\ipho}^2
,
\end{equation}
where $E_0$ is the equilibrium fixed-ions energy,
  $z_{\ipho}$ is a particular phonon coordinate 
  and $\allz$ denotes the ensemble of all of these coordinates.
Taking the lattice dynamics into account,
  the phonon eigenstates are those of the decoupled harmonic oscillators:
\begin{align}
  \chi^{\alln}(\allz)
    = \prod_{\ipho} \chi^{n_\ipho}(z_\ipho)
,
\end{align}
where $\alln$ denotes the ensemble of all the phonon occupation numbers.
The total energy in this state is 
\begin{align}
  E[\alln] = E_0 + \sum_{\ipho} \freqi \big[ n_\ipho + \tfrac{1}{2} \big]
.
\end{align}

The expression for a particular eigenvalue at finite temperature is
  given by the derivative of $F=E-TS$, the Helmholtz free energy
  with respect to an electronic occupation number $\feig$,
  which reduces to \cite{Ponce2014}
\begin{equation} \label{eigHelmholtz}
  \eig(T) = \frac{\del F}{\del\feig}
          = \eigo +
            \sum_{\ipho}  \frac{\del \freqi}{\del f_{\ieig}}
                  \big[ n_{\ipho} (T) + \tfrac{1}{2} \big]
.
\end{equation}
This expression should be compared with the electron-phonon self-energy
  in the adiabatic approximation (Eq.~\eqref{DWSE} and \eqref{staticFanSE}.)
The individual phonon contributions to the self-energy are proportional to
  $\del \freqi / \del f_{\ieig}$, which we call the
  electron-phonon coupling energies (EPCE).
Using Brook's theorem~\cite{Ponce2014}, the EPCEs are given by the second-order
  derivatives of an eigenvalue with respect to a phonon coordinate:
\begin{equation} \label{EPCEharmonic}
   \frac{\del \freqi}{\del f_{\ieig}} = 
   \frac{\del \eigi}{\del n_{\ipho}} = 
    \frac{1}{2\freqi}
    \frac{\partial^2}{\partial z_{\ipho}^2}
    \ \varepsilon_{\ieig} \big[ z_{\ipho} \big]
    \Big\vert_{z_{\ipho}=0}
,
\end{equation}
where $\varepsilon_{\ieig} \big[ z_{\ipho} \big]$ is an
  electronic energy computed with all atoms displaced by a length $z_{\ipho}$
  along the phonon displacement vector $U^{\ipho}_{\iat\icart}$.
This expression does not rely on the rigid-ion approximation,
  but requires a supercell calculation to account for the phonon wavevector.
Within the validity of the rigid-ion approximation,
  it should reproduce the results of the static DFPT scheme.
Both of these frameworks are developed within the
  harmonic approximation, since the total energy and the electronic
  eigenvalues are expanded up to second order in the phonon perturbations.
Equivalently, the harmonic approximation can be defined as the assumption
  that the electronic eigenvalues vary quadratically
  with a phonon coordinate $z_{\ipho}$.

In order to relax the harmonic approximation on the electronic energies,
  we cast the free energy $F=k_B T \ln Z$
  in terms of the canonical partition function $Z=\Tr \ e^{-\beta H}$,
  which is a trace over both the electronic and the atomic
  degrees of freedom.
Resolving the trace over the electron coordinates,
  the expression for a temperature-dependent eigenvalue reads
\begin{equation}
  \eig (T) = \idallz \frac{e^{-\beta E[\allz]}}{Z_I} \eig [\allz]
,
\end{equation}
where $Z_I=\idallz \ e^{-\beta E[\allz]}$
  is the partition function of the atoms only, and $\eig [\allz]$
  is the eigenvalue computed in some frozen-phonon configuration.
This formulation is reminiscent of the 
  path-integral molecular dynamics approach \cite{Ramirez2006, Ramirez2008},
  with the difference that a configuration is specified in terms of
  phonon coordinates rather than atomic positions in real space.
It retains the adiabatic approximation,
  since the atomic motion does not induce electronic transitions,
  but leaves the electrons in their evolving states.

We now use the crystal phonon structure
  and perform the harmonic approximation on the total energy only, writing
\begin{equation} \label{EPCEallorders}
  \eig(T) = \sum_{\alln} \frac{e^{-\beta E[\alln]}}{Z_I}
            \idallz \normallchi \eig [\allz]
.
\end{equation}
We are assuming here that the set of phonon wavefunctions
  $\chi^{n_\ipho}(z_\ipho)$ and frequencies $\freqi$ computed
  from second-order perturbation theory are good eigenfunctions of the system.
That is to say that the total energy is quadratic
  along the computed phonon modes even if the eigenvalues are not.
Equation~\eqref{EPCEallorders} now includes all high-order diagrams
  that may contribute to the self-energy,
  such as those depicted in Fig.~\ref{highorder}.

Finally, we perform the \emph{independent phonon approximation} and write
\begin{equation} \label{EPCEanharmonic}
  \eig (T) = \eigo +
             \sum_{\ipho} \sum_{n_\ipho}
             s^{n_{\ipho}}
             \int d z_\ipho \big\vert \chi^{n_\ipho}(z_\ipho) \big\vert^2 
             \Big( \eig [z_\ipho] - \eigo \Big)
,
\end{equation}
where $s^{n_{\ipho}}=e^{-\beta  \freqi n_\ipho}/
                    \sum_{n_\ipho'} e^{-\beta \freqi n_\ipho'}$.
In doing so we disregard the cross-terms contributions between different
  phonons modes.
This ansatz restricts the high-order diagrams to those containing a single
  phonon mode, which may interact multiple times with the electrons.
These additional diagrams come from the anharmonicity of the eigenvalues
  appearing in the integrant of Eq.~\eqref{EPCEanharmonic},
  as illustrated in Fig.~\ref{fig:EeAnharmonicity}.
One can verify that if the eigenvalues vary quadratically with the phonon
  displacements, then Eq.~\eqref{eigHelmholtz} is recovered.
Otherwise, Eq.~\eqref{EPCEanharmonic} defines effective EPCEs for each
  phonon mode which include the anharmonic effects.

\subsection*{Results and discussion}

We compute the EPCEs by frozen-phonon calculations,
  using the phonon displacement vectors obtained from DFPT.
For the harmonic approximation, Eq.~\eqref{EPCEharmonic}
  is evaluated with atomic displacements of about $10^{-3}\AA$,
  while the anharmonic effects are included by evaluating
  Eq.~\eqref{EPCEanharmonic} with $20$ displacements up to $\sim0.3\AA$,
  which corresponds to about 4 units of a typical phonon
  average displacement $1/\sqrt{\freqi}$.

\begin{figure}
  \includegraphics[width=1.0\linewidth]{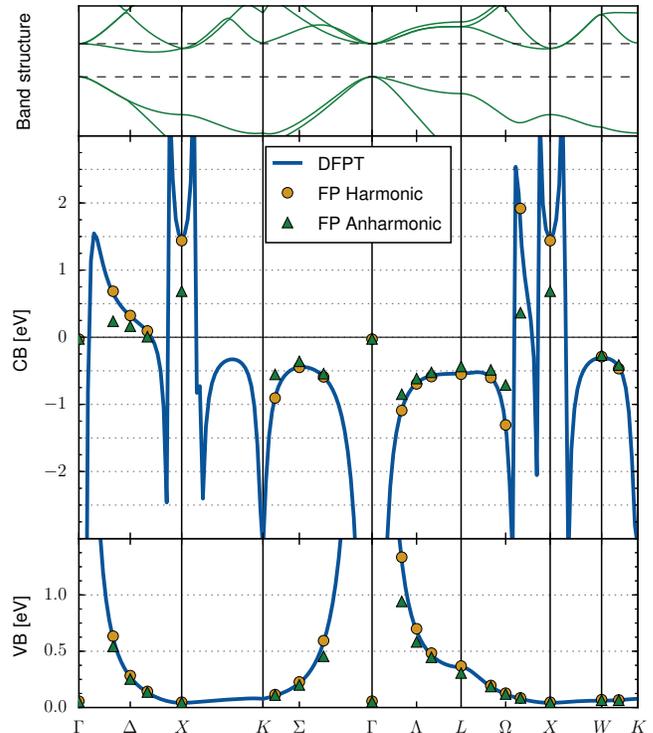}
  \caption{\label{fig:EPCEBS}
    Upper:
      The band structure of diamond, in eV.
      The dashed lines show the energies of the VB and CB states.
    Middle and lower:
      Electron-phonon coupling energies (EPCE), in eV for the CB state (middle)
      and the VB state (lower), computed with various methods.
    The blue line is the DFPT calculation, the yellow discs are
      the frozen-phonon method in the harmonic approximation, and the green
      triangles are the frozen-phonon method including anharmonic effects.
    A divergence is observed in the EPCEs of the CB state when a phonon
      wavevector couples this state to an other one with close energy,
    while the EPCE of both VB and CB states show a broad diverging peak
      at the center of the Brillouin zone.
      }
\end{figure}

The EPCEs are shown in Fig.~\ref{fig:EPCEBS} through the Brillouin zone
  of diamond.
The spiky structure of the EPCEs of the first conduction band at $\Gamma$
  results from this state not being at the bottom of the band.
Consequently, when a phonon wavevector connects the state at $\Gamma$
  to an other state with close energy, a divergence occurs in the EPCEs.
A divergence also occurs for phonon wavevectors near $\Gamma$,
  for both the VB and CB states, but these divergences integrate
  to a finite value when the density of phonon modes is taken into account.
The EPCEs computed with the frozen-phonon method in the harmonic approximation
  are in close agreement with the DFPT results, indicating that the
  rigid-ion approximation holds.
However, when the full dependence of the eigenvalues
  on the phonon displacements is taken into account,
  the anharmonicity of the eigenvalues tend to reduce the EPCEs,
  with respect to the harmonic approximation.

\begin{figure}
  \includegraphics[width=1.0\linewidth]{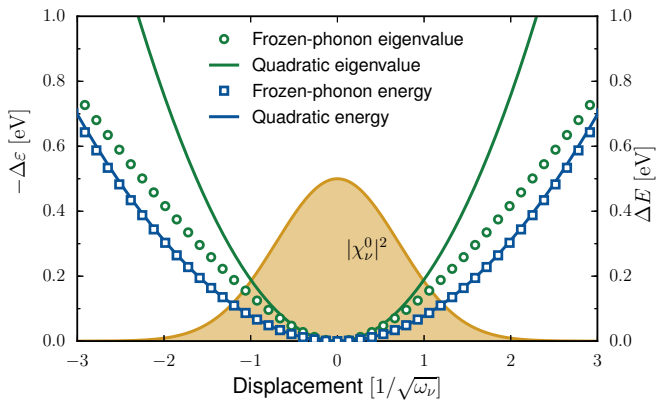}
  \caption{\label{fig:EeAnharmonicity}
    Dependence of the CB eigenvalue (green, circles)
      and the total energy (blue, squares)
      on a phonon displacement for the mode $\Omega_4$ of diamond.
    The circles and squares are the actual frozen-phonon calculations,
      the solid lines correspond to the harmonic approximation,
      and the filled curve is the phonon wavefunction.
      }
\end{figure}

This is exemplified on Fig.~\ref{fig:EeAnharmonicity} with the mode $\Omega_4$
  (the fourth mode with wavevectors $\Omega=(L+X)/2$).
In the harmonic approximation, it contributes $-869$~meV to the CB  EPCE
  at this q-point.
The eigenvalue however departs from the quadratic behavior
  with the phonon displacement, reducing the coupling energy to $-383$~meV.
On the other hand, the total energy follows closely the quadratic curve,
  indicating that this displacement is a genuine phonon mode.
This tendency is observed near all divergent points of the Brillouin zone
  and near the zone center.
The second-order perturbation theory is thus insufficient to treat
  the effect of those strongly coupling modes on the electronic states.

Table~\ref{tab:anharmonic} reports the ZPR computed on a
  $4\times4\times4$ q-point grid with the various static schemes.
Again, the total ZPR obtained with the harmonic frozen-phonon method
  and with DFPT are in good agreement.
The discrepancies can be attributed to the rigid-ion approximation.
When anharmonic effects are included, the total renormalization of the
  electronic energies is typically reduced
  compared to the harmonic approximation.
For the indirect band gap materials, the renormalization of the CB state
  is largely affected by the anharmonic effects,
  since they receive an important contribution
  from those strongly coupling modes at the Brillouin zone boundaries
  which are being attenuated.
The states at the band edges are being affected to various extends.
The valence band of LiF, which is especially flat,
  shows a strong anharmonicity in the ZPR coming from the modes near $\Gamma$,
  reducing the ZPR by about $60\%$.
In contrast, the conduction band of MgO, which is very dispersive,
  is only slightly affected by these effects.

Our results are obtained on a coarse q-point grid,
  limited by the scaling of the frozen-phonon method.
Whether the same conclusions apply to a converged q-point grid
  depends on the relative importance of strongly coupling modes,
  since they are responsible for anharmonic effects.
For the states lying at the top or at the bottom of their respective band,
  the ZPR increases monotonically with the number of q-points on a regular grid.
This is because the region of strongly coupling phonon modes near $\Gamma$
  gains in importance.
Thus, the anharmonic effects for these state are expected to grow
  as a finer q-point sampling is achieved.
For the states that are not at the extrema of their band,
  the convergence of the ZPR with q-points sampling is non-monotonic.
In these cases, we cannot make quantitative predictions
  for the anharmonic effects on a converged q-point sampling.
Nevertheless, the presence of strongly coupling modes
  near the Brillouin zone boundaries
  suggests an important anharmonic contribution to the ZPR.

\begin{table}
  \caption{
    Zero-point renormalization of the band gap (in eV)
      within the adiabatic approximation,
      obtained whith the static DFPT,
      with the frozen-phonon method in the harmonic approximation (FPH),
      and with the frozen-phonon method including anharmonic effects (FPA).
    \label{tab:anharmonic}}
  \begin{ruledtabular}
  \begin{tabular}{l l r r r r r}
    &  & Static DFPT
    & FPH & FPA\\
    \tline
    C &    VB &   0.115 &   0.119 &   0.107\\
      &    CB &  -0.320 &  -0.321 &  -0.214\\
      &   Gap &  -0.436 &  -0.439 &  -0.320\\
   BN &    VB &   0.120 &   0.133 &   0.108\\
      &    CB &  -0.193 &  -0.198 &  -0.154\\
      &   Gap &  -0.313 &  -0.331 &  -0.262\\
  MgO &    VB &   0.110 &   0.118 &   0.070\\
      &    CB &  -0.081 &  -0.078 &  -0.084\\
      &   Gap &  -0.191 &  -0.196 &  -0.154\\
  LiF &    VB &   0.445 &   0.431 &   0.168\\
      &    CB &  -0.130 &  -0.122 &  -0.113\\
      &   Gap &  -0.575 &  -0.553 &  -0.281\\
  \end{tabular}
  \end{ruledtabular}
\end{table}

\section{Conclusion}

The dynamical DFPT scheme allowed us to compute the frequency-dependent
  electron-phonon coupling self-energy.
Our calculations yield a renormalization factor
  ranging from $0.6$ to $1.0$.
This renormalization factor
  is important to obtain correct quasiparticle energies,
  but has been laregly overlooked in the literature.

The spectral function reveals distinctive features
  of the quasiparticle band structure.
In the indirect band gap materials (diamond and BN),
  the conduction bands undergo intra-band scattering processes,
  which broaden the spectral function at $\Gamma$.  
In the direct band gap materials with flat valence bands (LiF and MgO),
  these processes even generate satellite peaks
  below the valence bands.

The broadening can be obtained from the imaginary part of the self-energy,
  but one has to use a dynamical theory to do so.
Not only are the phonon frequencies necessary to impose
  energy conservation in the scattering process, but the imaginary part
  of the self-energy must be evaluated at the \emph{renormalized} eigenvalues,
  in order to compute properly the quasiparticle broadening.

Finally, we explored anharmonic effects using frozen-phonon calculations.
The anharmonicity in the eigenvalue dependence on the atomic displacements
  occurs even if the phonon modes are correctly described by the
  second-order perturbation theory.
This effect tend to decrease the contribution of the strongly coupling
  phonon modes, reducing the ZPR of certain states by as much as $60\%$
  with respect to the static AHC theory.

Our results indicate that high-order electron-phonon coupling terms
  bring an important contribution to the self-energy and the ZPR.
Our methodology however includes a partial summation
  of the high-order terms and treats the perturbations statically.
A theory that would include all high-order terms in a dynamical way
  cannot be tested at present,
  but could be eventually addressed with quantum Monte-Carlo approaches.

\begin{acknowledgments}
G. Antonius and M. C\^ot\'e thank the NSERC, the FRQNT and the RQMP for
  the financial support, and Calcul Qu\'ebec for the computational ressources.
This work was supported by the FRS-FNRS through a FRIA fellowship
  (S. Ponc\'e).
\end{acknowledgments}

\FloatBarrier

\appendix

\section{Imaginary parameter and convergence properties} \label{app:eta}

\begin{figure}
\includegraphics[width=1.0\linewidth]%
  {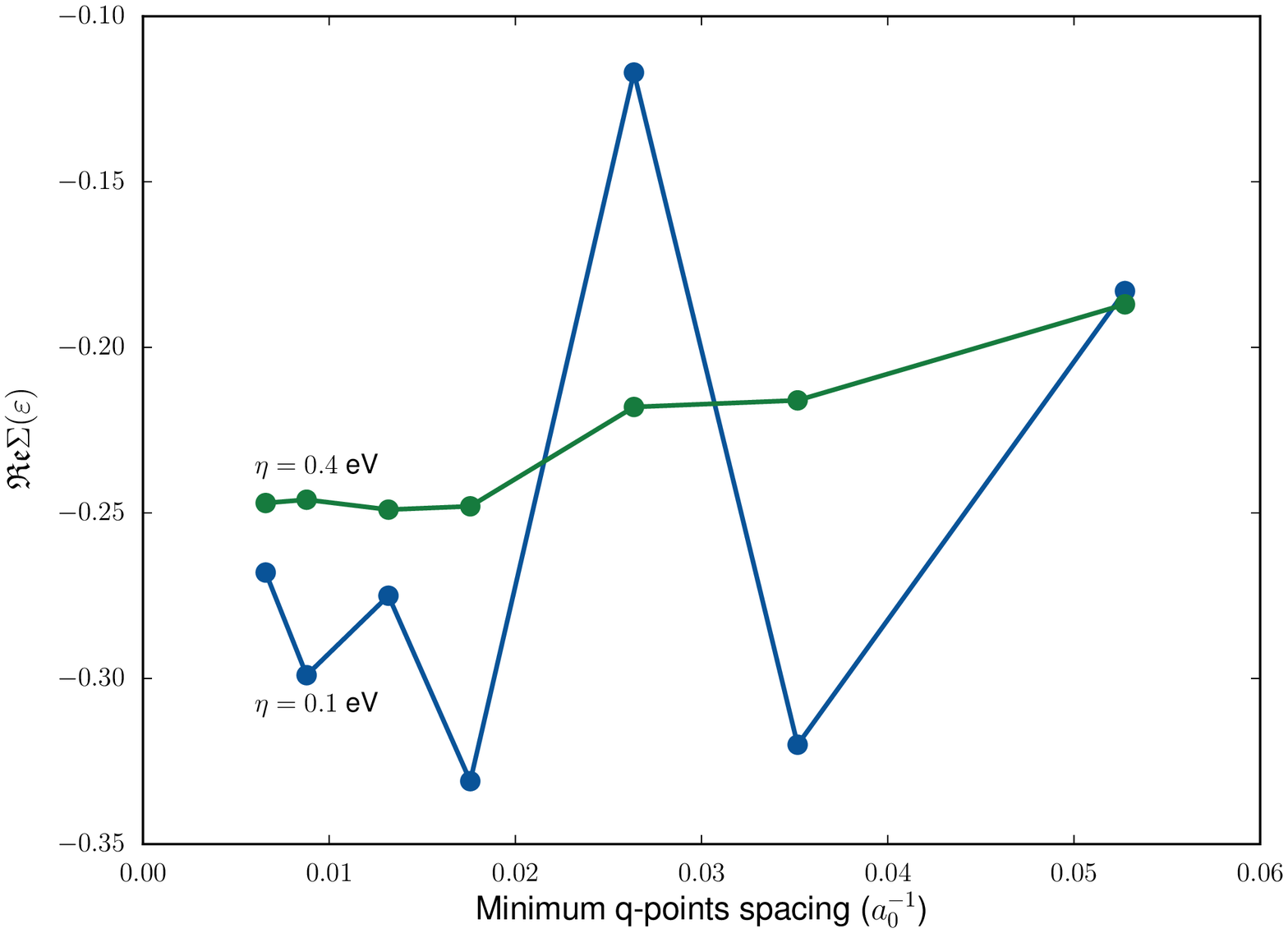}\\
\includegraphics[width=1.0\linewidth]%
  {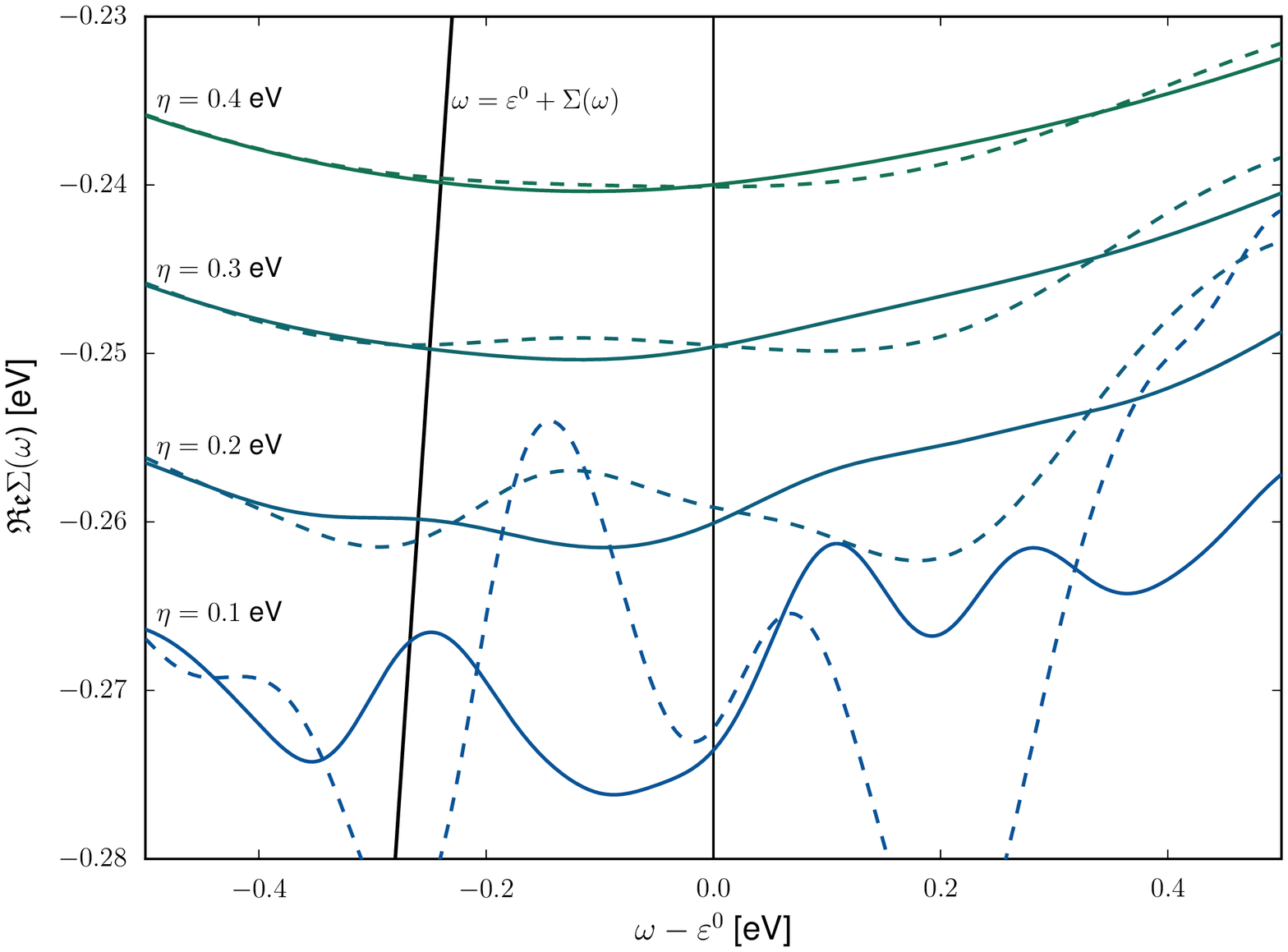}
  \caption{\label{fig:DFPTconv}
    Upper:
      Convergence of the self-energy at the renormalized eigenvalue
      for the CB state of diamond,
      as a function of q-points spacing,
      with various imaginary parameters.
    Lower:
      Frequency dependence of the self-energy for the CB state of diamond,
      with various imaginary parameters.
      The solid lines are obtained on a $32\times32\times32$ q-point grid,
      and the dashed lines are obtained on a $24\times24\times24$ q-point grid,
      corresponding to the two left-most points on the upper pannel graph.
           }
\end{figure}

In order to compute the ZPR, one has to sample the phonon wavevectors
  in the Brillouin zone, either through a regular mesh,
  or with a random set of q-points.
At the same time, one has to select a value for the parameter $\eta$
  giving an imaginary part to the self-energy.
The choice of this parameter should in facts be addressed
  in conjunction with the q-points sampling.
When the static DFPT method is used, the numerical value assigned to $\eta$
  is usually on the order of typical phonon frequencies ($\sim 0.1$~eV)
  to account for their omission.
Otherwise, if a too small value of $\eta$ is used,
  it is not even clear that the self-energy will converge
  to a finite value\cite{ponce_temperature_2015}.
In a dynamical scheme, one should in principle aim for a vanishing
  value for $\eta$.
While it is expected that the self-energy elements should converge to a
  finite value as $\eta \rightarrow 0$, tuning the value of $\eta$
  conveniently eases the convergence with the number of q-points,
  as shown in Fig.~\ref{fig:DFPTconv}.

Moreover, using a too small value of $\eta$ can compromise
  the frequency dependence of the self-energy,
  as shown on Fig.~\ref{fig:DFPTconv} for the CB state of diamond.
Even for the most converged q-point grid, the self-energy computed
  with $\eta=0.1$~eV shows rapid variations with $\omega$.
These variations are even larger for a smaller q-point grid,
  and in these cases, the solution of $\omega=\Sigma(\omega+\varepsilon^0)$
  could certainly not be estimated by linearizing the self-energy near the
  bare eigenenergy.
The self-energy becomes a perfectly smooth function of $\omega$
  when $\eta=0.4$~eV.

We use the following criterion to determine the value of $\eta$.
Consider the contribution of two neighboring q-points $\vq$ and $\vq'$
  to the self-energy of the electronic state $\vk n$.
The contribution of a particular electron band $n$ and phonon branch $m$
  will have terms proportional to
$\big[(\omega - \varepsilon^0_{\vk+\vq n} \pm \omega_{\vq m} + i\eta)^{-1} +
 (\omega - \varepsilon^0_{\vk+\vq' n} \pm \omega_{\vq' m} + i\eta)^{-1}\big]$,
  assuming that the matrix elements in the numerator of the self-energy
  does not change between $\vq$ and $\vq'$.
If the value of $\eta$ is vanishingly small,
  the spectral function will exhibit distinct peaks at
  $\omega=\varepsilon^0_{\vk+\vq n}\pm\omega_{\vq m}$ and
  $\omega=\varepsilon^0_{\vk+\vq' n}\pm\omega_{\vq' m}$,
  which will be an artifact of the q-points sampling.
The separation of those peaks comes mainly from the dispersion
  of the electronic energies, which is more important
  than that of the phonon frequencies.
Simple analysis shows that these peaks can be made undistinguishable
  by setting ${\eta = \sqrt{3} \Delta \varepsilon / 2}$ where
  $\Delta \varepsilon=\varepsilon^0_{\vk+\vq n}-\varepsilon^0_{\vk+\vq' n}$.
Hence, for a given q-point mesh, we compute the largest $\Delta \varepsilon$
  between neighboring q-points, within the bands being corrected,
  and use it to deduce $\eta$.
The values of $\eta$ obtained for our most converged q-point grid
  are: $0.2$~eV for the VB state of diamond, $0.4$~eV for the CB states
  of diamond and BN, and $0.1$~eV for all the other VB and CB states.
We verified that the broadening of the CB states of diamond and BN
  was insensitive to the choice of this parameter.

\end{document}